\documentclass[trackchanges]{aastex7}

\begin{document}

\title{Detection of Variability in Seyfert 2 Galaxies and Measurement of the Optical Scattering Region Size}

\author[0009-0002-4279-9925]{Natalie Kovacevic}
\affiliation{Homer L.\ Dodge Department of Physics and Astronomy,
University of Oklahoma, Norman, OK 73019, USA}
\email{natalie.kovacevic@ou.edu}

\author[0000-0001-9203-2808]{Xinyu Dai}
\affil{Homer L.\ Dodge Department of Physics and Astronomy,
University of Oklahoma, Norman, OK 73019, USA}
\email{xdai@ou.edu}

\author[0000-0002-7720-3418]{Heechan Yuk}
\affil{Homer L.\ Dodge Department of Physics and Astronomy,
University of Oklahoma, Norman, OK 73019, USA}
\email{}

\author{Emilia Jaervelae}
\affiliation{Department of Physics and Astronomy, Texas Tech University, Box 41051, Lubbock, TX, 79409-1051, USA}
\affiliation{Homer L.\ Dodge Department of Physics and Astronomy,
University of Oklahoma, Norman, OK 73019, USA}
\email{}

\author[0000-0001-8920-0073]{Tingfeng Yi}
\affiliation{Key Laboratory of Colleges and Universities in Yunnan Province for High-energy Astrophysics, Department of Physics, Yunnan Normal University, Kunming 650500, China}
\email{}

\author[0000-0001-5661-7155]{Patrick J. Vallely}
\altaffiliation{NSF Graduate Research Fellow}
\affiliation{Department of Astronomy, The Ohio State University, 140 West 18th Avenue, Columbus, OH 43210, USA}
\email{}

\author[0000-0003-4631-1149]{Benjamin J. Shappee}
\affiliation{Institute for Astronomy, University of Hawai\`{}i at Manoa, 2680 Woodlawn Dr., Honolulu, HI 96822, USA}
\email{}

\author[0000-0001-8973-5051]{Francesco Shankar}
\affiliation{School of Physics and Astronomy, University of Southampton, Highfield, Southampton 1BJ, UK}
\email{}

\begin{abstract}
One main theme of the Unification Model of active galactic nuclei is that there is an obscuring torus structure blocking the direct view of the central engine for Seyfert 2 galaxies.
Here, we present the detection of long-term optical variability for a sample of nearby Seyfert 2s.
We found that Seyfert 2s exhibit a relatively low but significant level of variability beyond the constant fluxes established by galaxies over month to year time scales.  
The variability is also detected in the structure functions of Seyfert 2s.
Assuming a simple variability suppression model by the scattering region and dilution due to host starlight, where the region smooths the unobscured photon packets from the central engine as the light scatters over the torus, we estimate AGN scattering region sizes by matching the variability amplitudes of Seyfert 1s to 2s.  Our measured scattering sizes are largely consistent with the torus size measured using their emission properties, suggesting that the scattering region is of a similar size as the torus. Our results pave the way towards variability as a powerful and independent test for AGN unification models.

\end{abstract}

\keywords{\uat{Active galactic nuclei}{16} --- \uat{Seyfert galaxies}{1447} --- \uat{Supermassive black holes}{1663} --- \uat{Galaxy accretion disks}{562}}

\section{Introduction} \label{intro}
Recent studies revealed that most massive galaxies house a supermassive black hole (SMBH) at their center \citep{kormendy_1995ARA&A..33..581K}. A fraction of these SMBHs will actively accrete the surrounding material, producing a luminous light source visible across the electromagnetic spectrum. This type of SMBH is often referred to as an active galactic nucleus (AGN).

\begin{figure}[ht]
\centering
\includegraphics[width=0.60\textwidth,trim={1cm 0cm 0cm 1cm}]{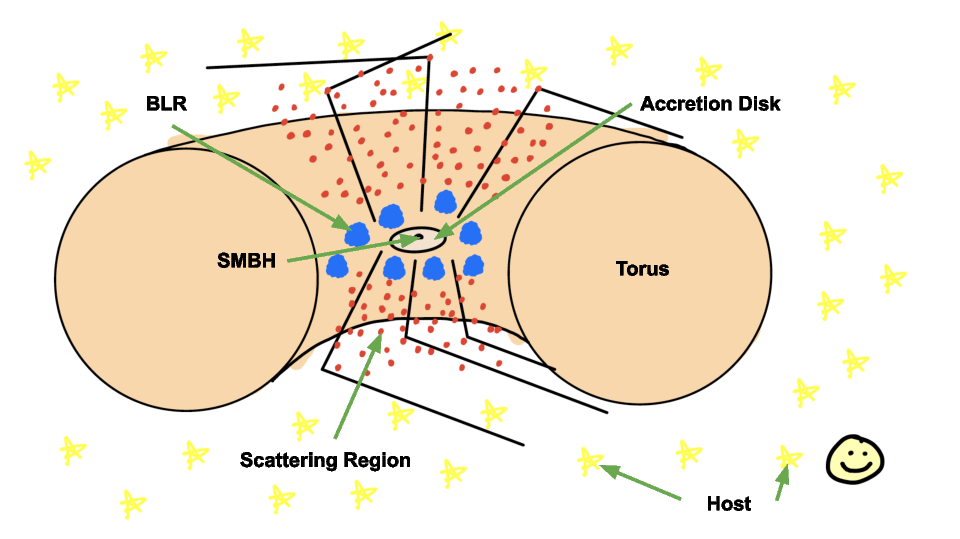}
\includegraphics[width=0.37\textwidth,trim={1cm 0cm 0cm 1cm}]{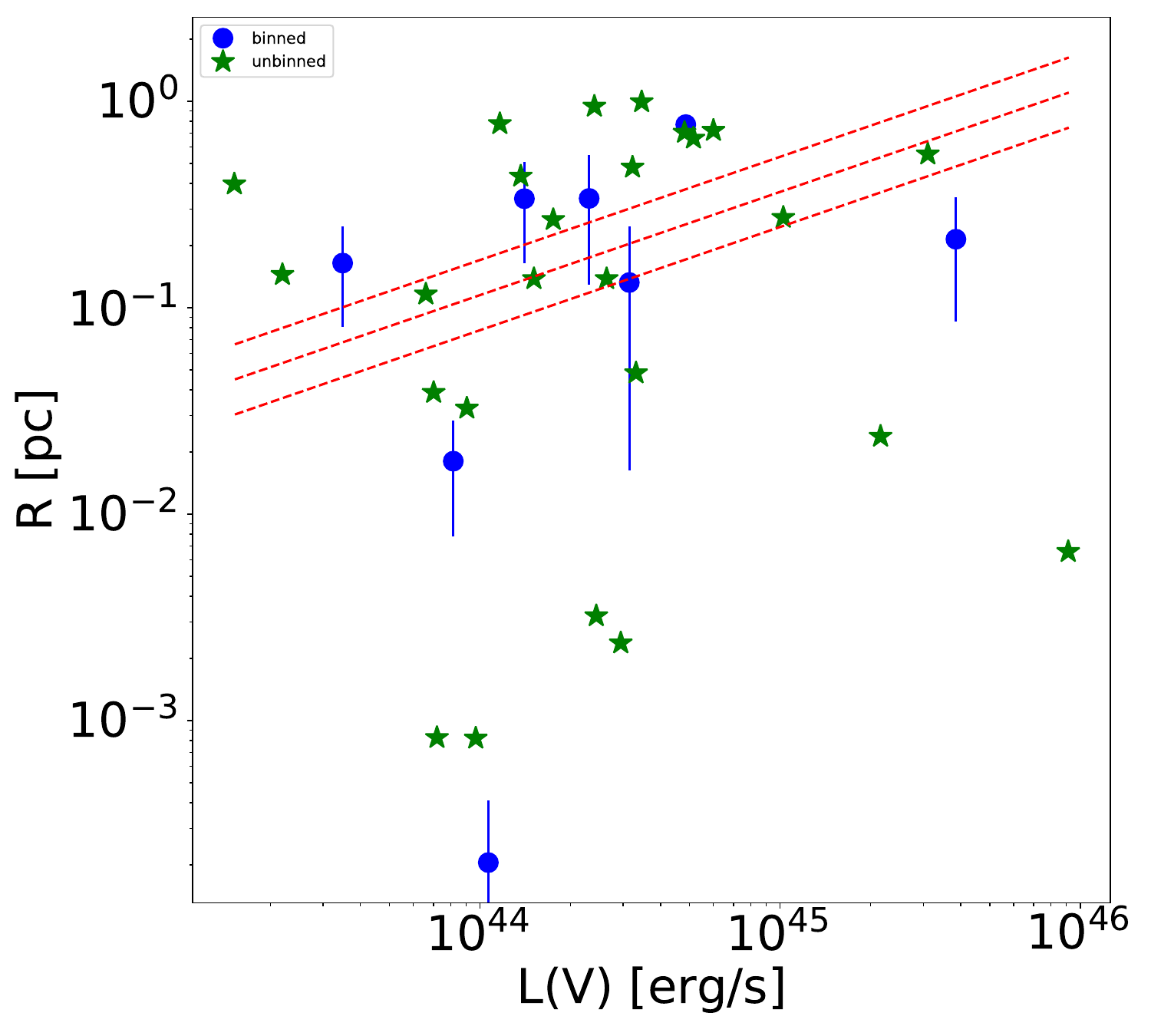}
\caption{Left panel: Schematic illustration for the interpretation of Sy2 variability. The central engine includes the black hole and the accretion disk, which is surrounded by an obscuring torus. The broad line regions are inside the torus, while the narrow line region and stars from the host are outside. Light is scattered into our line of sight via a scattering region (red dots) that dilutes the variability of Sy2s. This scattering region is expected to be close to the inner edge of the torus. The blue clouds represent the HBLR of many Sy2s and the yellow stars represent the contribution of host stellar light. Observers of Sy2s receive a combination of scattering emission from the central engine with suppressed AGN variability and starlight from the host galaxy. Right panel: The estimated sizes of the scattering region from our Sy1 sample by matching the Sy1 and 2 variability with a variability suppressing model, showing both the binned results in blue circles
and unbinned in green stars. The dashed lines are the reverberation mapping torus relation \citet{Kishimoto_2007A&A...476..713K} with upper and lower bounds indicating scatter of the relation \citet{Yang_2020ApJ...900...58Y}. A dilution fraction of 70\% was added to the smoothed light curves of Sy1s. The binned measurements of the scattering region using the variability method are consistent with the reverberation mapping torus size.}\label{interp}
\end{figure}

Seyfert galaxies \citep{Seyfert_1943ApJ....97...28S} were among the first galaxies discovered to contain AGNs in the 1940s through their distinct broad emission line features in their spectra. The SMBH within Seyfert galaxies can range in mass from 10$^6$ to 10$^9$ M$_\odot$ and tend to be less luminous than QSOs. Over the next half-century after the discovery, many AGNs classes were discovered and astronomers have developed the Unification model to unify different AGN classes under coherent schemes. AGN classes are divided according to their viewing angle as illustrated in Figure \ref{interp} left (\citealt{Antonucci_1993ARA&A..31..473A, Urry_1995PASP..107..803U}). Seyfert 1s (Sy1) are considered to have a direct line of sight to the central engine, resulting in spectra containing both broad and narrow emission lines, produced from the small scale broad line region (BLR) surrounding the accretion disk and the larger scale narrow line region (NLR) extending past the dust torus. Seyfert 2s (Sy2) have an obscured view of the BLR and central engine due to the dusty torus that surrounds the BLR, resulting in only narrow lines present in the spectra. One supporting piece of evidence in the development of the unification model is 
from modeling the integrated X-ray background (XRB). The existence of obscured AGNs, those that have an obscured view of the central engine, is required in conjunction with unobscured AGNs to accurately model the observed XRB spectrum \citep{Madau_10.1093/mnras/270.1.L17}. This is due to the obscuring material absorbing the lower energy X-rays while emitting a reprocessed emission component, allowing the higher energy X-rays to pass, producing harder X-ray spectra compared to unobscured AGNs contributing significantly to the hard X-ray portion of the XRB spectrum \citep{Treister_2005ApJ...630..115T}. However, a key piece of evidence for the unification model is from polarization results. From analyzing the polarized optical spectra of Sy2s, broad lines can be detected from scattered emission originating from the BLR, surrounded by the dust torus often referred to as the hidden BLR (HBLR). The most extensive polarization measurements are performed on the archetypal Sy2 NGC 1068 \citep{Antonucci_1985ApJ...297..621A}. Currently, spectral polarimetry measurements have been obtained for dozens of Seyfert 2s and a fraction of them exhibit a HBLR (\citealt{Wu_2011ApJ...730..121W, ichikawa_2015ApJ...803...57I, Ramos_2016MNRAS.461.1387R}). Besides the detections of HBLRs, polarization measurements also constrain the polarization angle close to 90 degrees relative to the jet axis, suggesting polar direction scattering. In addition, the polarization analyses also show that the optical continuum of Sy2s are mainly \citep{Antonucci_1985ApJ...297..621A} composed of the host stellar contribution, that is on average 70\%, and scattered emission from the central engine by material within the scattering region. The size of the scattering region is not well constrained, as it can be the torus itself, the inner edge of the torus, or the region within the inner edge of the torus. 

The variability of AGNs is a key feature observed across the electromagnetic spectrum, and the observed variability of AGNs is stochastic in nature, throughout all time scales. The mechanism responsible for AGN variability can have multiple physical origins for example X-ray reprocessing \citep{Krolic_1991ApJ...371..541K}, accretion
disk instabilities (\citealt{Kawaguchi_1998, Trèvese_2002}), star collisions \citep{courvoisier_1996A&A...308L..17C}, and supernova explosions in host galaxies or accretion
disk \citep{Aretxaga_10.1093/mnras/286.2.271}, as possible explanations. Galaxies lacking an active SMBH will exhibit no optical variability, with the observed flux being dominated by the non variable ensemble of stellar light. With Sy1 having a clear line of sight to the central engine, the optical variability of Sy1s has been studied countlessly from different time scales \citep{Webb_2000}. On the long time scale (years), most Sy1s exhibit flux variation (\citealt{deRuiter_1986A&AS...63...59D,Winkler_1992MNRAS.257..659W, winkler_1997MNRAS.292..273W, lopez_2023MNRAS.518.1531L}) while on the short time scale (weeks) this significantly decreases \citep{deRuiter_1986A&AS...63...59D}. If the unification model holds true, one would expect variability arising from Sy2s. However, due to the obscuration of the central engine of Sy2s only a few studies have been carried out on their variability, and Sy2s are generally found to exhibit negligible variability in the optical band in the short time scale (weeks) \citep{deRuiter_1986A&AS...63...59D} possibly due to the direct emission being largely absorbed and the scattered emission will smooth out the variations from different packets of photons, raising the question of whether Sy2s exhibit variability.

In this study, we present a comparison of the variability of Sy1, Sy2, and galaxy samples using two different all sky surveys at various time scales. Our results from the comparison show that Sy2 variability is detected with significant confidence. This was determined from comparing the normalized excess variance of the three samples at different time scales, further confirmed by the analysis of their structure functions. Using the comparison result, we calculated the size of the scattering region by utilizing a simple variability suppression model based on the unification model (Figure \ref{interp} left) on our Sy1 sample showing sizes consistent with torus size studies, constraining the scattering region to be consistent with the size of the inner torus radius (Figure \ref{interp} right). Throughout the paper we assumed cosmological parameters of H$_0$ = 72 km s$^{-1}$Mpc$^{-1}$, $\Omega_m$ = 0.3, and $\Omega_\Lambda$ = 0.7.

\section{Sample and Data Calibrations} \label{s_DC}
\subsection{\textit{Swift}-BAT 9-Month Survey and 2MASS Extended Source Catalog} \label{sample}

Our sample of Sy1s and Sy2s are derived from the \textit{Swift}-BAT 9-Month Survey \citep{Tueller_2008ApJ...681..113T}. This is a hard X-ray all sky survey on the Neil Gehrels Swift Observatory \citep{Gehrels_2004ApJ...611.1005G} using the Burst Alert Telescope \citep{Barthelmy_2005SSRv..120..143B}. From the parent sample we excluded sources that were classified as Blazar, BL Lac, LINER, and galaxy, limiting our sample to the Sy1, Sy2 and intermediate Seyfert types. Seyfert types were then grouped as obscured or unobscured based on the relative strength or absence of a broad H$\beta$ component. Sy1, Sy1.2 and Sy1.5 are grouped as unobscured (Sy1s) and Sy1.8, Sy1.9 and Sy2 are grouped as obscured (Sy2s) \citep{Osterbrock_1981ApJ...249..462O}. We also excluded sources that were identified to contain a radio jet according to their radio loudness $F_{5\textrm{GHz}}$/$F_B$ $>$ 10 \citep{Chiaberge_2011MNRAS.416..917C}, the inclusion of a relativistic jet provides an additional source of variability. Lastly, known Changing-Look (CL) AGNs were excluded from the final sample. The final Seyfert sample included 42 Sy1s and 38 Sy2s.

The parent sample for the galaxies are taken from the Two Micron All Sky Survey (2MASS) \citep{skrutskie_2006AJ....131.1163S} extended source catalog \citep{Jarrett_2000AJ....119.2498J}. The parent sample contains over 1 million galaxies, several cuts were performed to obtain our final sample. The first cut was a magnitude cut of K $<$ 11.34. Next was a cut on the galactic latitude $|b|$ $>$ 15$^{\circ}$ to avoid contamination from the galactic plane. Following this cut, ASAS-SN light curves were extracted for the remaining 12,435 galaxies. Based on the ASAS-SN light curves a second magnitude cut was performed 11.5 $<$ V $<$ 16.5. This will provide Galaxies that match the V-band apparent magnitudes of our Sy1 and Sy2 sample such that an unbiased comparison can be performed. The final cut was performed on the flux uncertainties within the ASAS-SN light curves. The measurement of variability is related to the flux uncertainties presented in the light curves, and to avoid a biased comparison, the galaxies in this sample need comparable ($<$ 15\%) flux uncertainties to our Sy2 sample. This final cut brought the galaxy sample to 43 galaxies.

\subsection{Extraction of ASAS-SN and TESS Light Curves} \label{extraction}
The data used for this study were obtained from the All-Sky Survey for SuperNovae (ASAS-SN) and the Transiting Exoplanet Survey Satellite (TESS), probing time scales over 5 orders of magnitude from hours to a decade.

\begin{figure}[ht]
\centering
\includegraphics[width=0.49\textwidth,trim={0cm 0cm 0cm 1cm}]{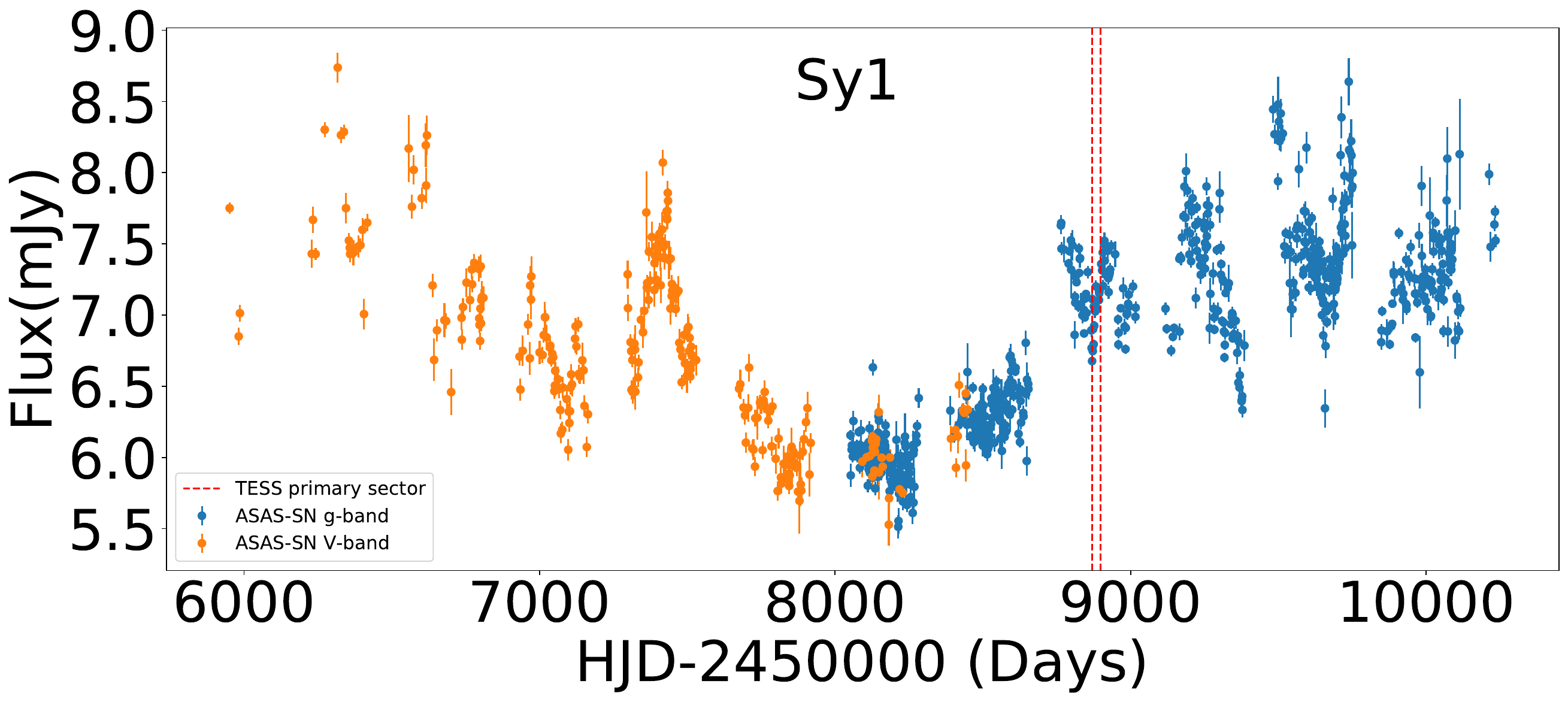}
\includegraphics[width=0.49\textwidth,trim={0cm 0cm 0cm 1cm}]{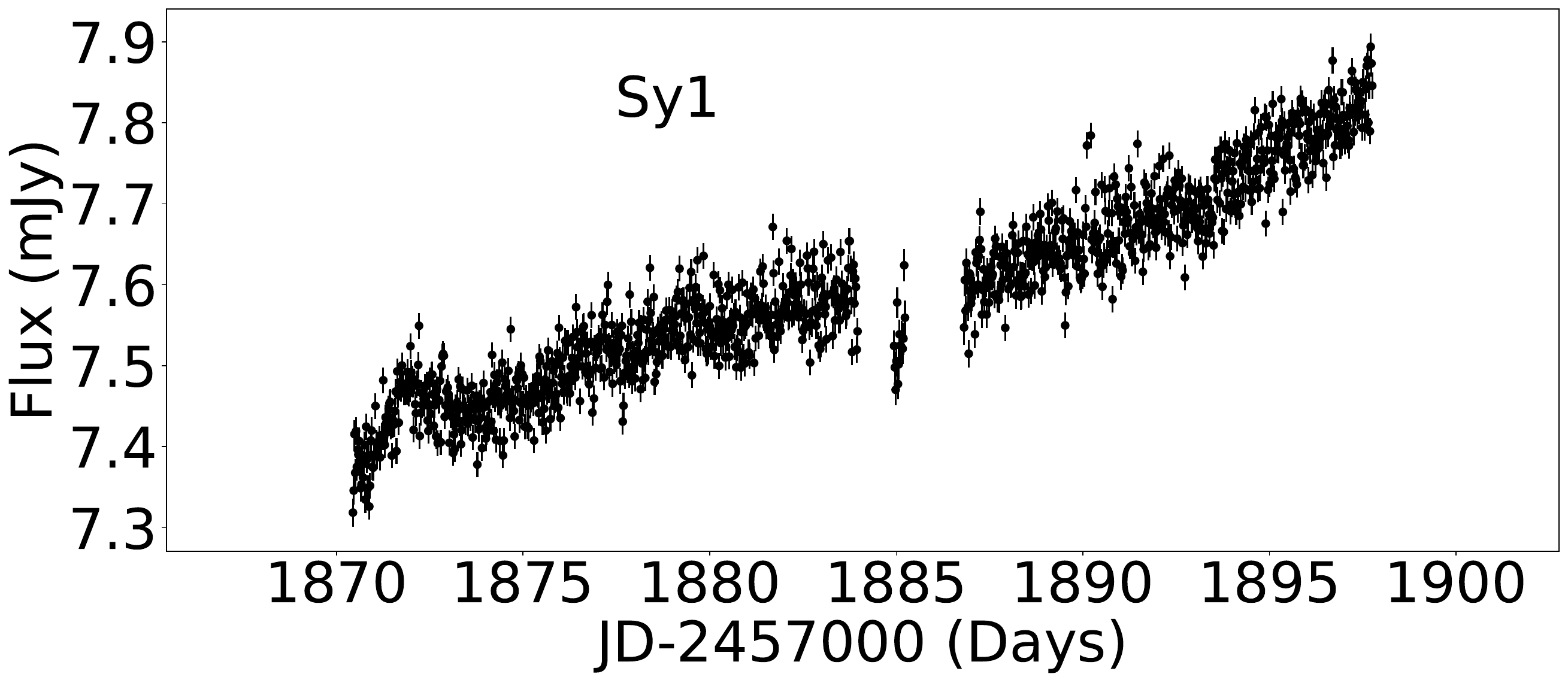}
\caption{Left panel: ASAS-SN light curve of Sy1 MCG+04-22-042 as an example of our sample. The red dotted lines indicate the time frame during which the TESS light curve occurs. Right panel: TESS light curve. ASAS-SN and TESS light curves show significant variability.}\label{sy1}
\end{figure}

\begin{figure}[ht]
\centering
\includegraphics[width=0.46\textwidth,trim={2cm 0cm 0cm 3.6cm}]{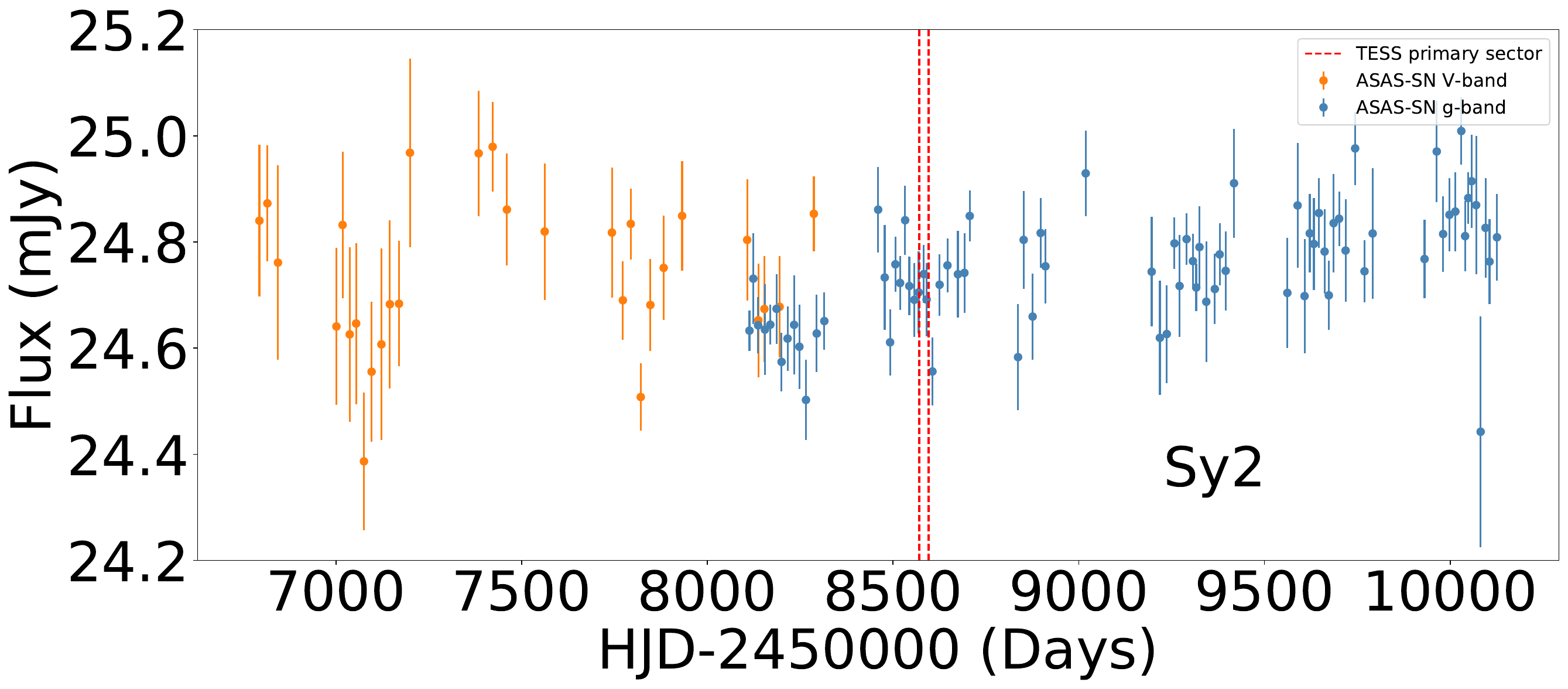}
\includegraphics[width=0.49\textwidth,trim={0cm 0cm 2cm 3.6cm}]{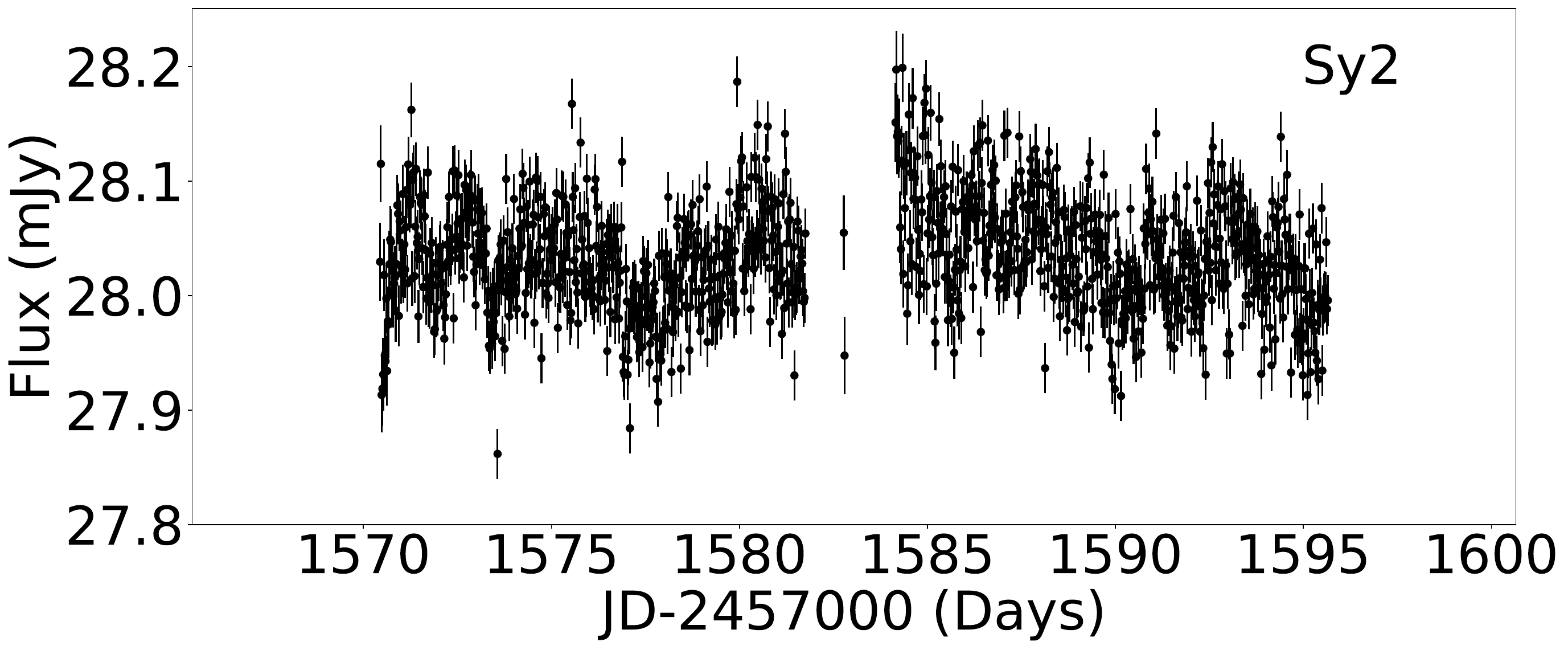}
\caption{Left panel: ASAS-SN light curve of the Sy2 NGC 4507, an example of our sample binned by 30 days. Right panel: TESS light curve. The ASAS-SN light curve shows suppressed AGN variability beyond a constant flux established by the galaxy light curves. The TESS light curve shows fluxes consistent with the constant flux established by the galaxy light curves}\label{sy2}
\end{figure}

\begin{figure}[ht]
\centering
\includegraphics[width=0.47\textwidth,trim={2cm 0cm 0cm 2cm}]{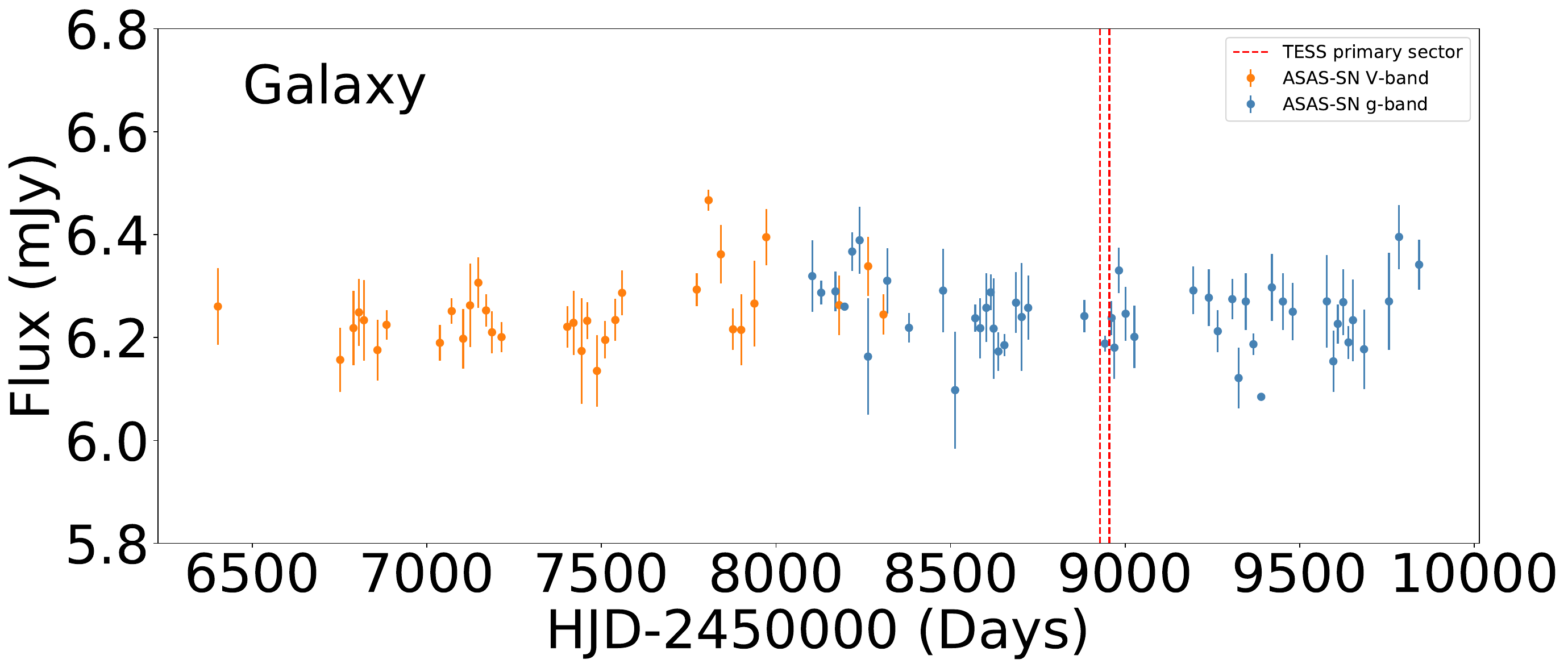}
\includegraphics[width=0.47\textwidth,trim={0cm 0cm 2cm 2cm}]{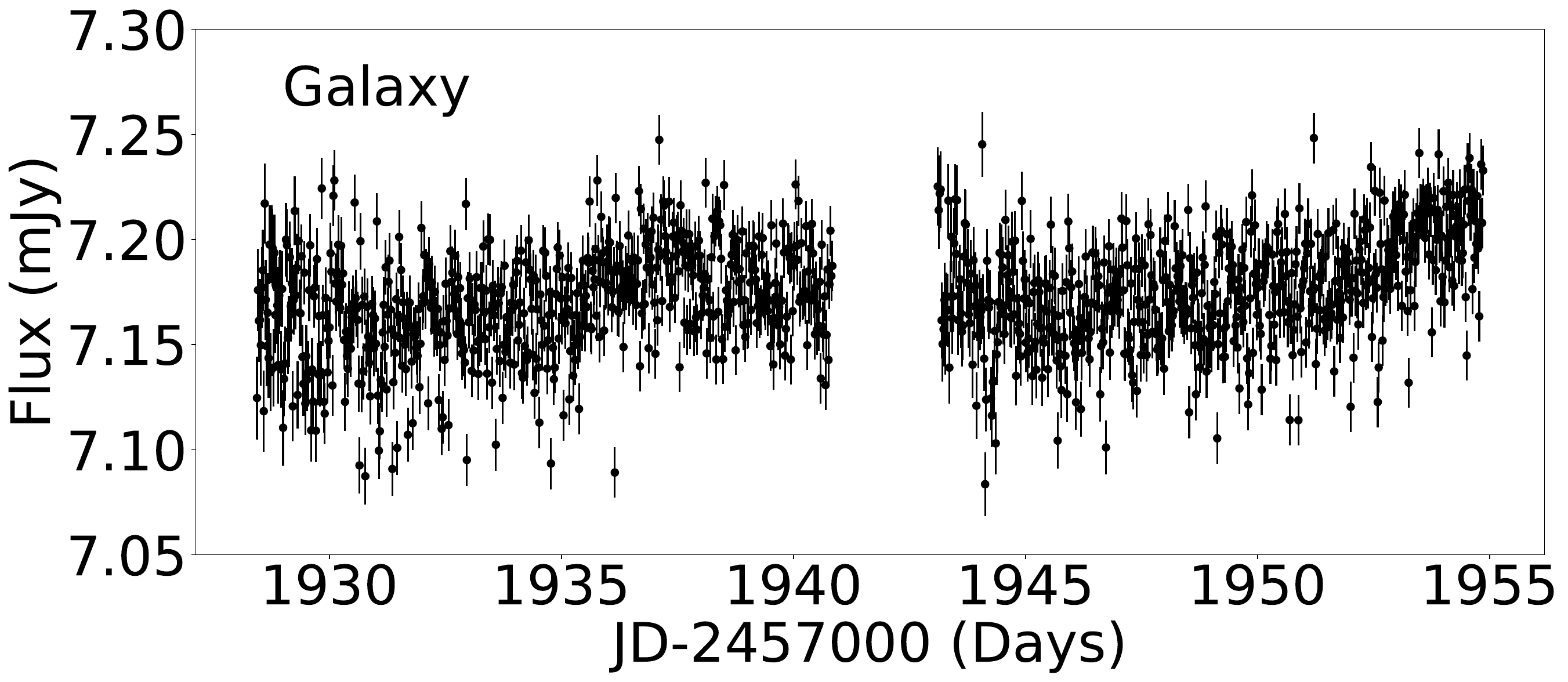}
\caption{Left panel: ASAS-SN light curve of the Galaxy UGC 9684 as an example of our sample binned by 30 days. Right panel: TESS light curve. Both galaxy light curves are expected to be constants, and they show the systematic and measurement uncertainties of the surveys.}\label{gal}
\end{figure}

ASAS-SN, consists of a suite of telescopes that have been observing the entire visible night sky for over a decade (\citealt{Shappee_2014,hart_2023arXiv230403791H}), making ASAS-SN a useful tool for studying variability at longer time scales \citep{Yuk_2022}.The ASAS-SN light curves were extracted by using image subtraction and aperture photometry, calibrated against an ensemble of stars in the observing field (\citealt{Jayashinghe_10.1093/mnras/stz444,jayasinghe_10.1093/mnras/sty838}). The V-band and g-band light curves are normalized by utilizing the mean flux of each band. 
Examples of a typical ASAS-SN Sy1, Sy2 and galaxy light curve can be seen in the left panels of Figures \ref{sy1}, \ref{sy2}, and \ref{gal} spanning roughly 3250 days. The first portion of the light curve is from the V-band filter, while the latter portion utilizes the g-band filter; each portion spans 1500 days and 2050 days, respectively, with an approximate 400-day overlap. Visually, the variability is present in the Sy1 light curve, while the Sy2 light curve shows suppressed variability that is beyond the constant flux presented in the galaxy light curve, which can be revealed by the numerical analysis of the light curves.

The precision and high cadence of TESS, ranging from 10 to 30 minutes, coupled with being an all-sky survey, makes TESS ideal for studying the variability of AGNs on the short time scale (\citealt{treiber_10.1093/mnras/stad2530,Yuk_2025A&A...698A.105Y}).We extracted the TESS light curves using their full frame images (FFIs) from the primary (sectors 1--26) and first extended mission (sectors 27--55). An automated image subtraction pipeline was applied during the TESS light curve extraction process \citep{vallely_2021MNRAS.500.5639V}. This pipeline utilizes four subtraction methods; the basic ISIS subtraction, median filtering, Gaussian smearing, and both the Gaussian smearing and filtering. To lessen the effect of CCD strap artifacts found in many TESS observations and based on visual inspection for systematic effects, the median filtering method was the most appropriate for further analysis. This method incorporates a background filtering that uses one-dimensional median filters which are applied along each axis of the CCD, following the initial ISIS subtraction. The FFIs used in this work were obtained from the Mikulski Archive for Space Telescopes (MAST)\footnote[1]{The MAST website is at https://doi.org/10.17909/0cp4-2j79}. Examples of a typical Sy1, Sy2 and galaxy TESS light curve can be seen in the right panels of Figures \ref{sy1}, \ref{sy2}, and \ref{gal} with each light curve comprising of two 13 day cycles and a one day cycle gap for a total light curve length of 27 days. The variability is visually seen in the Sy1 light curve; however, the Sy2 light curve is consistent with a constant flux established by the galaxy light curves.

\section{Analysis}

\subsection{Normalized Excess Variance} \label{NXS}

To quantify the intensity of variability in the case of AGN light curves, the normalized excess variance is often calculated (\citealt{vaughan_2003MNRAS.345.1271V,schleicher_2019Galax...7...62S}). Which takes the variance $S^2$ of the light curve excluding the uncertainties in the flux measurements $\sigma_{\textrm{err}}$, then normalized by the mean of the light curve $\langle{x}\rangle$. This can be seen in Equation \ref{eq1} below. 

\begin{equation}
\sigma_{\textrm{nxs}}^2 = \frac{S^2 - \langle{\sigma_{\textrm{err}}^2 }\rangle}{\langle{x}\rangle^2} \label{eq1}
\end{equation}
with the error of the normalized excess variance as,

\begin{equation}
\Delta{\sigma_{\textrm{nxs}}^2} = \sqrt{\Biggl( \sqrt{\frac{{2}}{N}}\frac{\langle{\sigma_{\textrm{err}}^2 }\rangle}{\langle{x}^2\rangle} \Biggr)^2 + \Biggl({\sqrt{\frac{\langle{\sigma_{\textrm{err}}^2 }\rangle}{N}}\frac{2F_{\textrm{var}}}{\overline{x}}}\Biggr)^2} \label{eq2}
\end{equation}
where $N$ is the number of data points and the square root of the normalized excess variance commonly called the fractional variability $F_{\textrm{var}}$ is shown below.

\begin{equation}
F_{\textrm{var}} = \sqrt{\sigma_{\textrm{nxs}}^2}\label{eq3}
\end{equation}

Most of our TESS light curves include a single sector of data for which the normalized excess variance was calculated, for the light curves that included multiple sectors, the normalized excess variance was calculated for the combined light curve.

\begin{figure}[ht]
\centering
\includegraphics[width=0.48\textwidth,trim={0cm 0cm 0cm 0cm}]{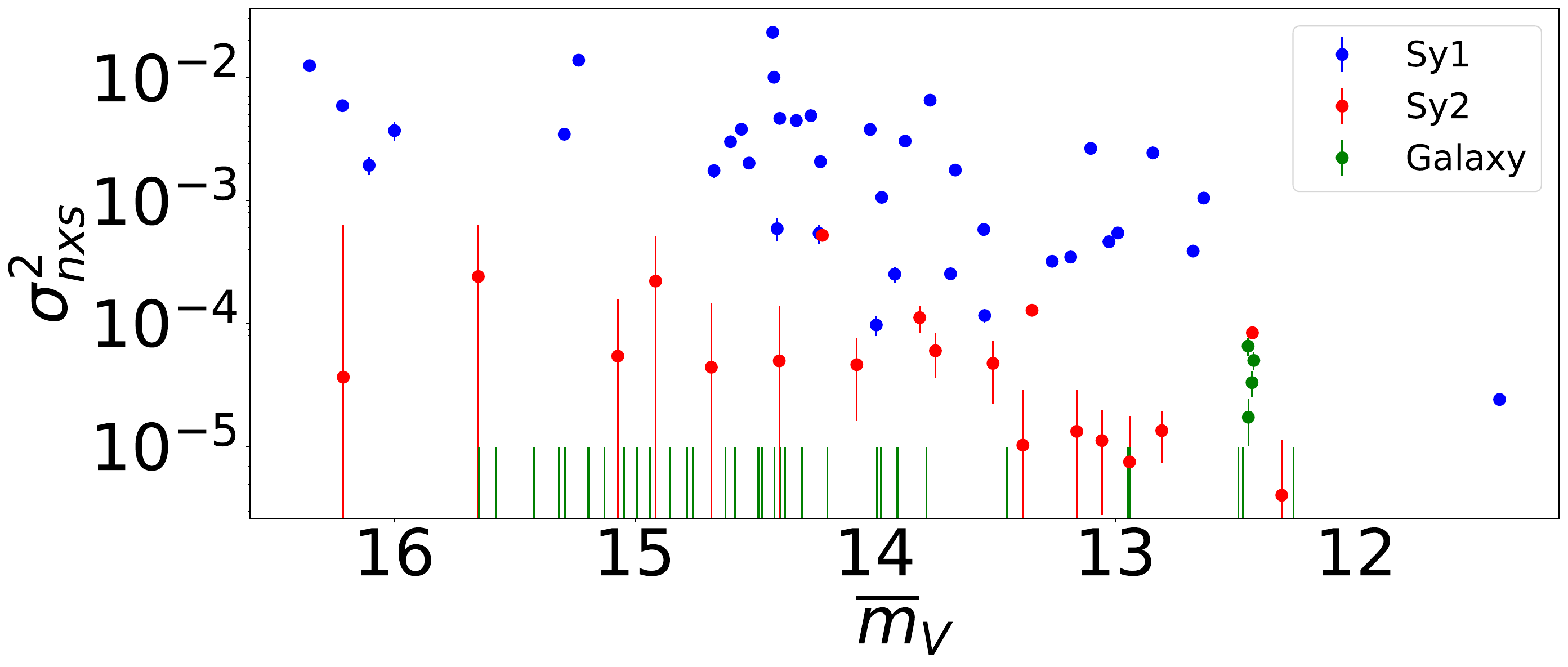}
\includegraphics[width=0.48\textwidth,trim={0cm 0cm 0cm 0cm}]{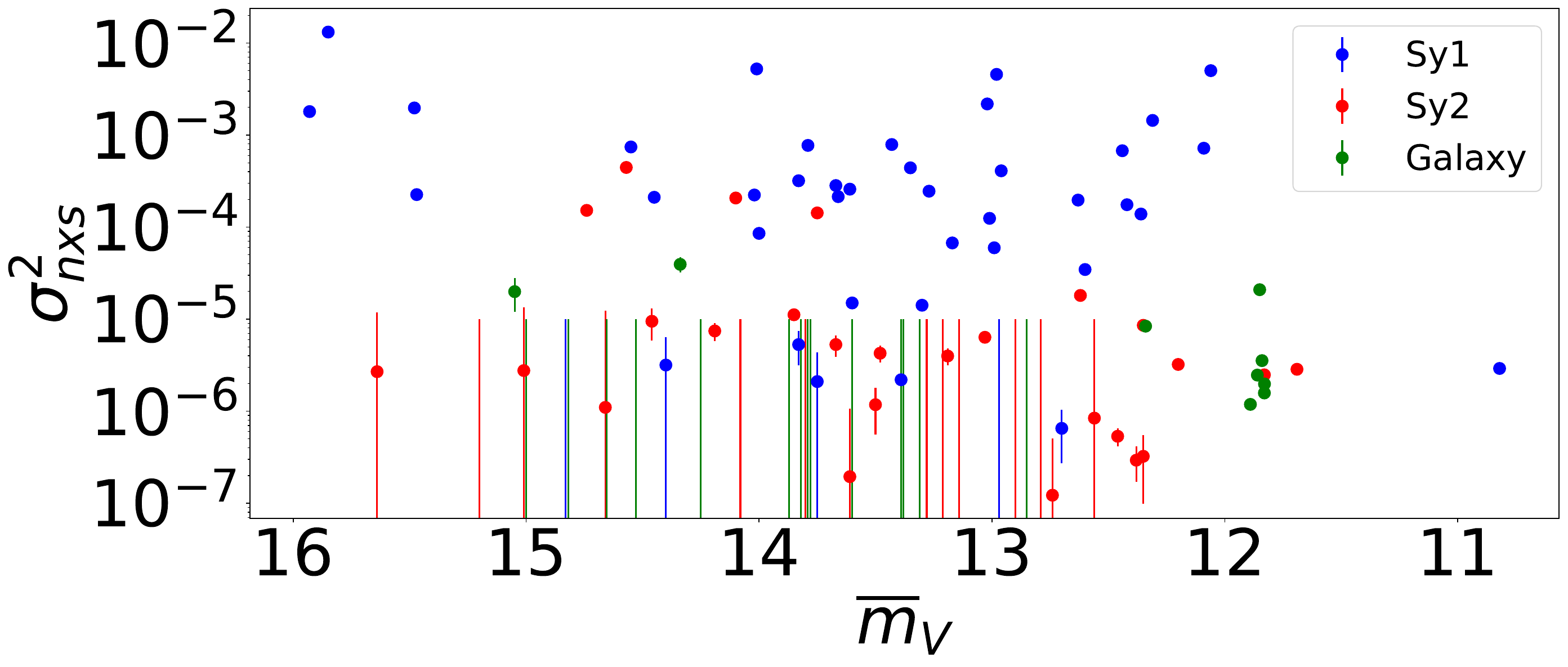}
\caption{Left: The relationship between the average apparent V-band magnitude and normalized excess variance for the ASAS-SN
light curves. After correcting for the underestimation of flux measurement uncertainties, in the long time scale ASAS-SN data the excess variance of the Sy2 sample sits higher than that of the galaxy sample, signifying a detection of Sy2 variability. Right: The relationship between the average apparent TESS magnitude and normalized excess variance for the TESS light curves.}\label{app_mag}
\end{figure}

We check for additional systematic uncertainties by examining the relationship between the median apparent magnitude and the normalized excess variance of the TESS and ASAS-SN data sets. This reveals an increasing trend in the normalized excess variance as the sources magnitudes get fainter. This trend was present in all three samples and was attributed to an underestimation of flux measurement uncertainties present in the light curves. To correct for this, a correction factor was applied first to our Sy2 sample that corrected the error term presented in Equation \ref{eq1} for a given apparent magnitude range such that the mean $\sigma_{\textrm{nxs}}^2$ in that range coincided with the lowest Sy2 $\sigma_{\textrm{nxs}}^2$ value. This correction eliminated the dependence of $\sigma_{\textrm{nxs}}^2$ with the apparent magnitude for the Sy2 sample. These correction factors ranging from 1 to 8.4 were then applied to the Sy1 and galaxy samples. Figure \ref{app_mag} displays the relationship between $\sigma_{\textrm{nxs}}^2$ and the apparent magnitude after all corrections were applied.  

Most of the ASAS-SN light curves in our sample include seasonal observational gaps. Although the data is roughly evenly sampled a bias correction is needed to account for these seasonal gaps. Using equation 11 described in \citet{Allevato_2013ApJ...771....9A} with a PSD slope of $\beta = 1$ \citet{Yuk_2025A&A...698A.105Y} no bias correction was applied as the correction factor is 1. The TESS data however, have a 1 day cycle gap with most light curves being evenly sampled. To correct for this cycle gap again using equation 11 in \cite{Allevato_2013ApJ...771....9A} and a PSD slope of $\beta = 2$ \citep{Yuk_2025A&A...698A.105Y}, a bias correction factor of 0.48 was applied.

\subsection{Structure Functions} \label{SF_section}
The structure function is a correlation-like method often used to analyze the variability time scale for AGNs \citep{emman_10.1111/j.1365-2966.2010.16328.x}. The structure function  is related to the variance of a light curve at large time scales. One common definition for the structure function presented in \citet{Macleod_2010ApJ...721.1014M} and \citet{szymom_2016ApJ...826..118K} is 

\begin{equation}
\textrm{SF}(\tau) = \sqrt{\langle{[m(t+\tau)-m(t)]^2\rangle}-\sigma_{\textrm{noise}}^2}\label{eq4}
\end{equation}

Where the structure function is calculated using a magnitude at time t $(m(t))$ and the time difference between data pairs ($\tau$). The noise contribution is defined as 

\begin{equation}
\sigma_{\textrm{noise}}^2 = \langle{\sigma_{\textrm{error}}^2(t)+\sigma_{\textrm{error}}^2(t+\tau)\rangle}\label{eq5}
\end{equation}

and the $\sigma_{\textrm{error}} (t)$ is the error on the measured magnitudes.
The structure function at very long time scales $t_{\infty}$ is related to the standard deviation of the light curve as
\begin{equation}
\textrm{SF}_{\infty} = \sqrt{2}\sigma\label{eq6}
\end{equation}

Examples of stacked and individual Sy1, Sy2, and galaxy structure functions are presented in Figures~\ref{sf_stacked} and \ref{sf}.
\section{Results}

\subsection{Detection of Variability in Seyfert 2s} \label{KS_test}

\begin{figure}[ht]
\centering
\includegraphics[width=0.4\textwidth,trim={0cm 0cm 0cm 0cm}]{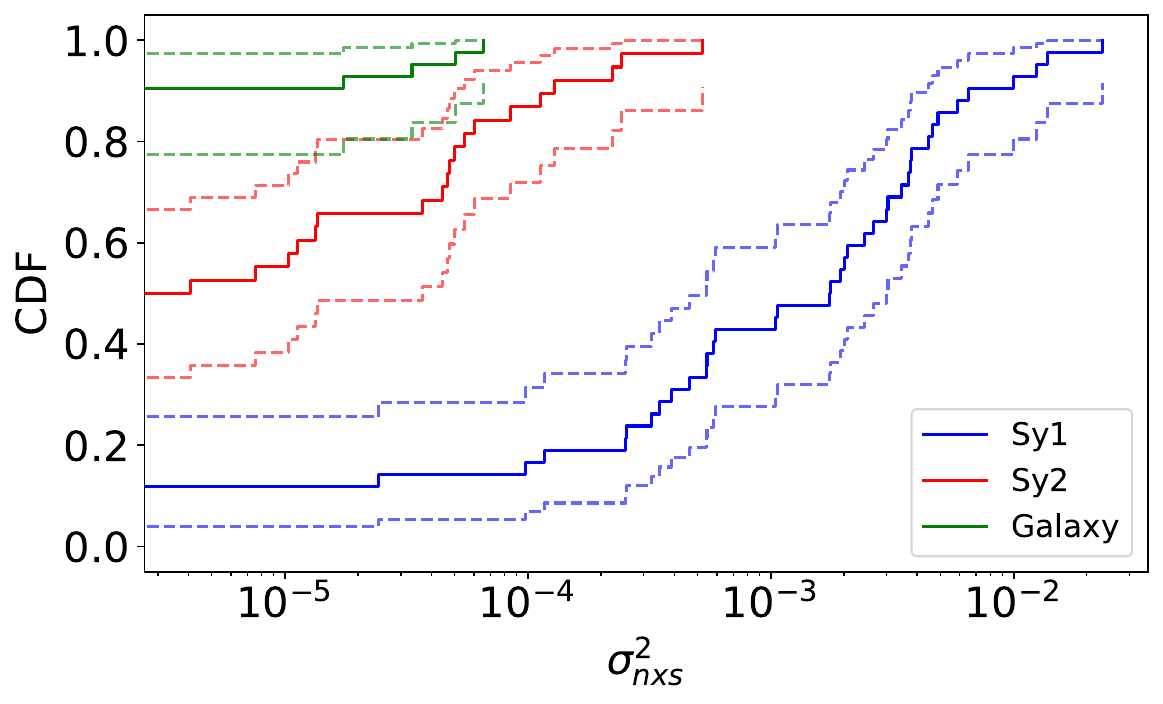}
\includegraphics[width=0.4\textwidth,trim={0cm 0cm 0cm 0cm}]{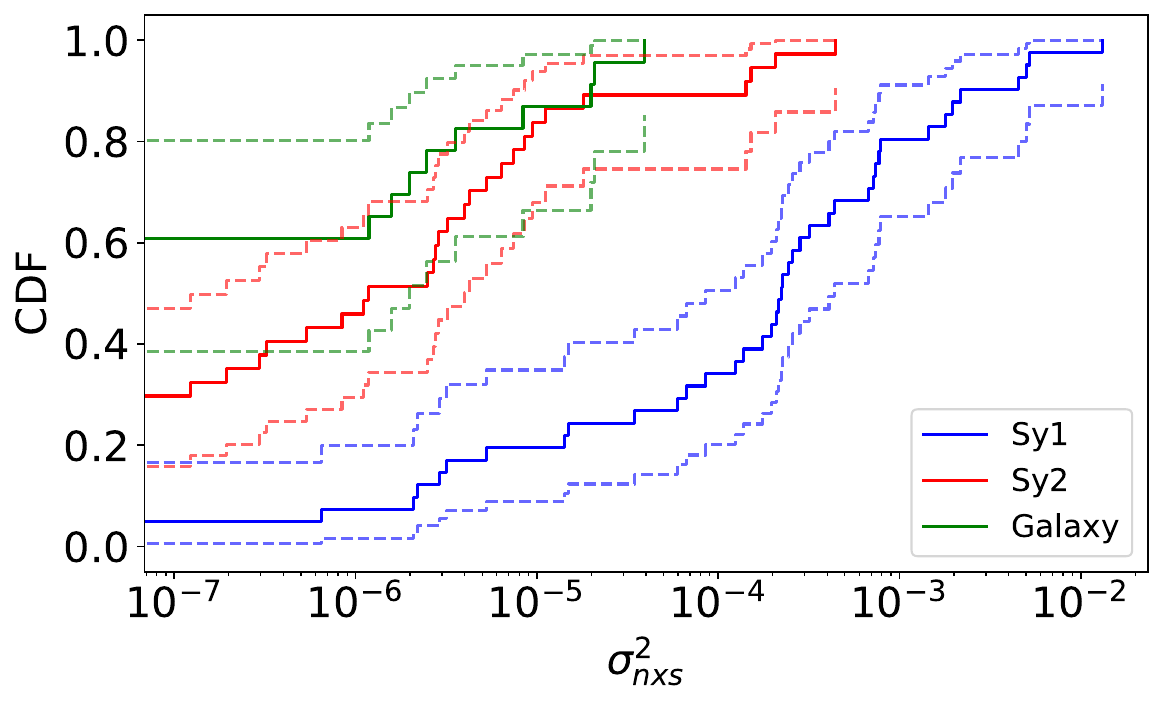}
\caption{The cumulative distribution function of the normalized excess variance of Sy1s and Sy2s, compared to the constant flux established by galaxy measurements from the ASAS-SN (Left) and TESS (Right) data sets. The dotted lines represent the Clopper-Pearson confidence bands. Visually from the ASAS-SN measurement plot, it is evident that Sy2s have variabilities different than that of galaxies with a p-value of $8.4\times10^{-4}$. The TESS data set, however, yields a p-value of 0.13, signifying there is no difference on the short time scale between galaxy and Sy2 light curves.}\label{KS}
\end{figure}

We measured the variability of the sample by calculating the excess variance, a noise removed variance of the light curve. To quantify the differences in the distribution of variabilities of the Sy1, Sy2, and galaxy samples a two-sample Kolmogorov-Smirnov (KS) test was performed on the normalized excess variance. The tests were performed on the sample of Sy2s and galaxies using the TESS and ASAS-SN data sets, yielding p-values of 0.13 and $8.4\times10^{-4}$ respectively. The result of the KS test on the TESS dataset shows that Sy2s and galaxies are not significantly different at the TESS time scales. 
The corresponding p-value using the ASAS-SN dataset (\textbf{p=$8.4\times10^{-4}$}) is well below the null-hypothesis threshold (p $<$ 0.01, \citet{Wasserstein02042016}), suggesting that Sy2 variability is detected on longer time scales of months to years. This is because galaxies should exhibit no variability due to the dominance of the non-variable ensemble stellar light in their light curves, acting as a benchmark for AGN variability. The cumulative distributions of $\sigma_{\textrm{nxs}}^2$ for the Sy1, Sy2, and galaxy samples using the ASAS-SN and TESS data sets are shown in Figure \ref{KS}. Confidence bands were constructed around all three samples using the point wise Clopper-Pearson interval \citep{clopper_10.1093/biomet/26.4.404} which gives the 95\% confidence interval for individual points. Using the ASAS-SN data the distributions for Sy2s and galaxies are visually distinguished with only a nonsignificant overlap in their confidence interval. Previously, KS-tests were performed on the Sy1 and Sy2 sample using the ASAS-SN and TESS data sets, yielding p-values of $1.4\times10^{-12}$ and $2.4\times10^{-7}$ respectively \citep{kovacevic_2025ApJ...985..177K}, establishing the variability difference between the Sy1 and Sy2 sample in both long and short time scales.

\begin{figure}[ht]
\centering
\includegraphics[width=0.4\textwidth,trim={2cm 0cm 2cm 0cm}]{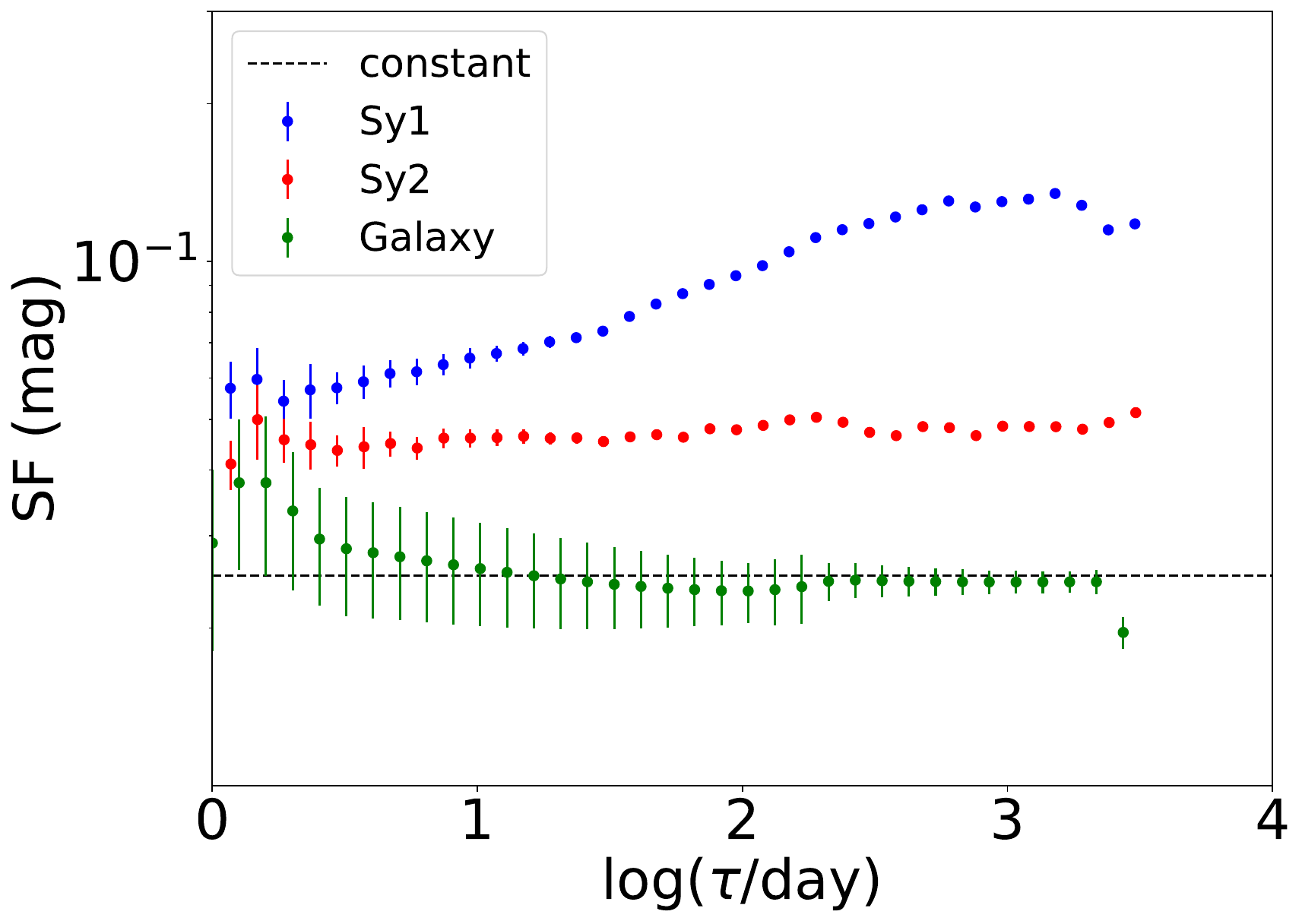}

\caption{Stacked Sy1, Sy2, and galaxy structure functions calculated using the ASAS-SN dataset.}\label{sf_stacked}
\end{figure}

\begin{figure}[ht]
\centering
\includegraphics[width=0.4\textwidth,trim={1cm 0cm 0cm 1.0cm}]{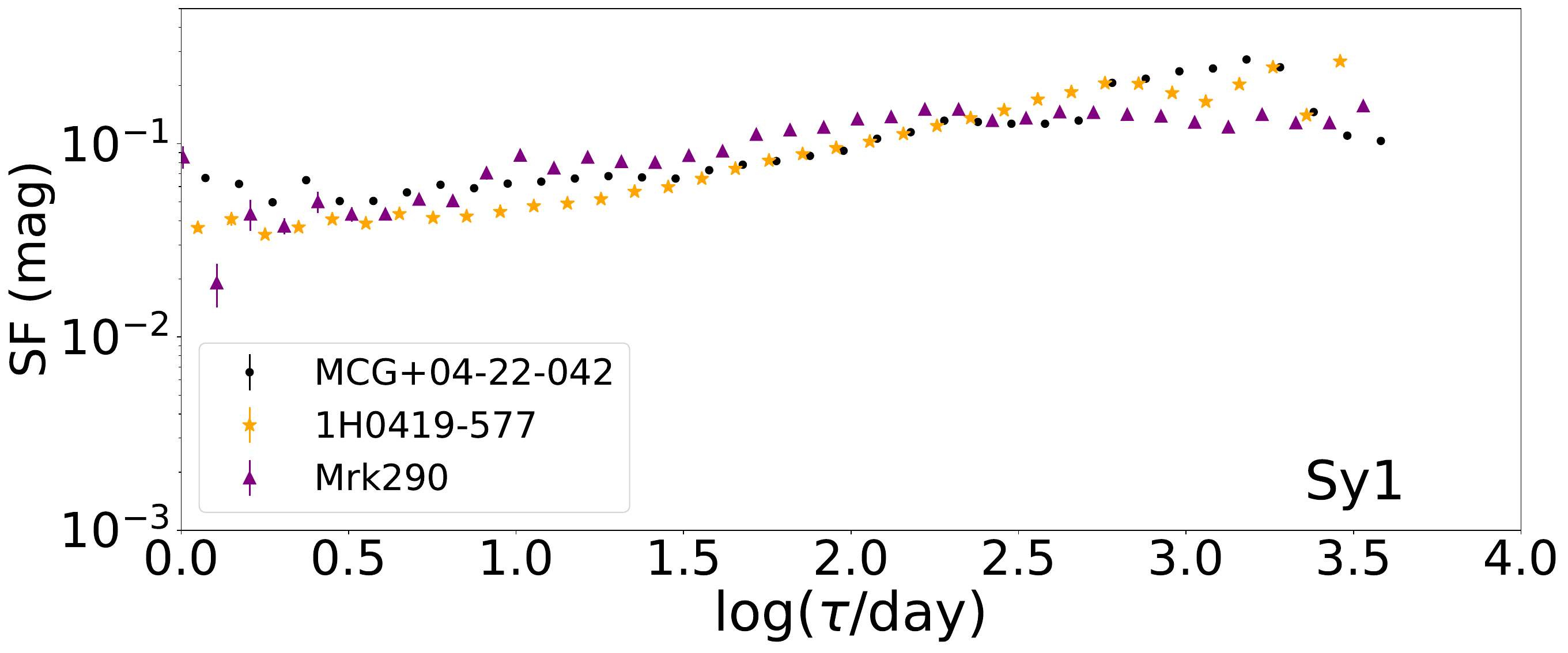}
\includegraphics[width=0.4\textwidth,trim={0cm 0cm 1cm 2.0cm}]{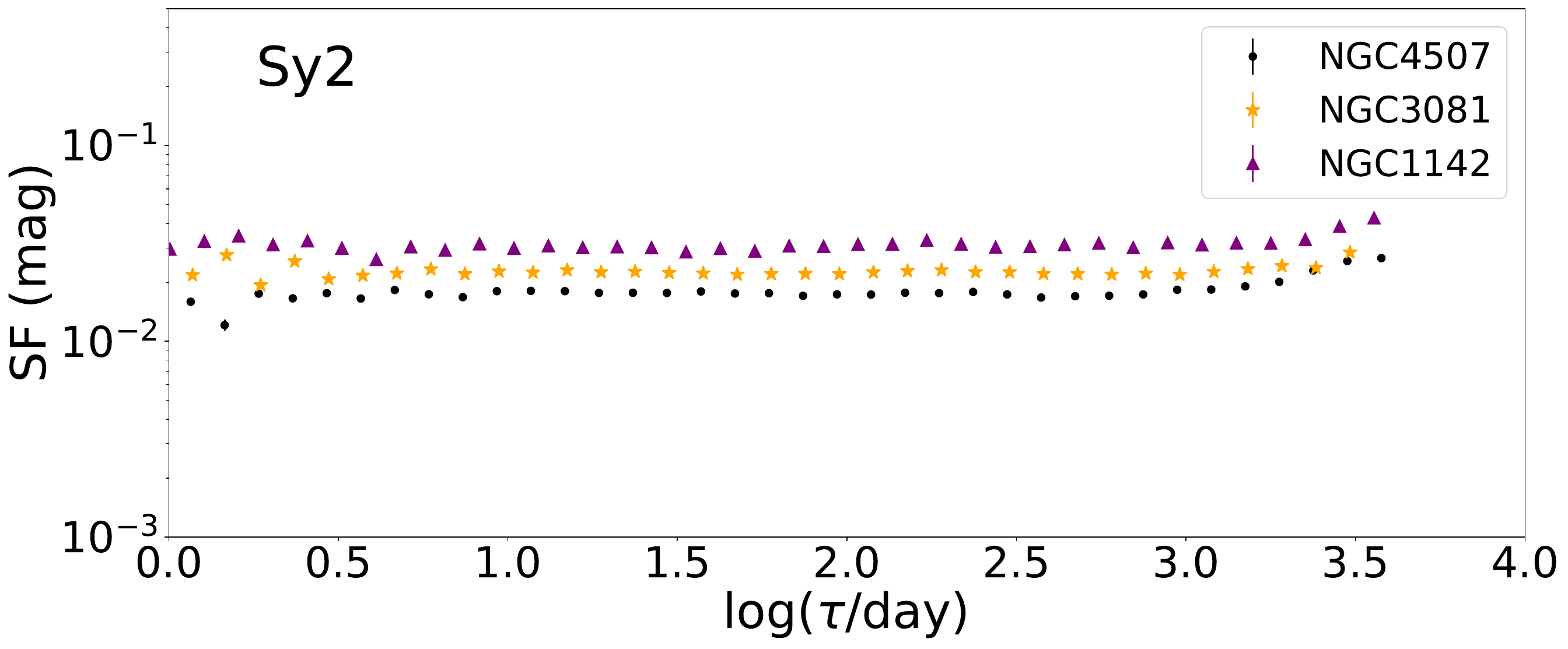}
\includegraphics[width=0.4\textwidth,trim={2cm 0cm 2cm 0cm}]{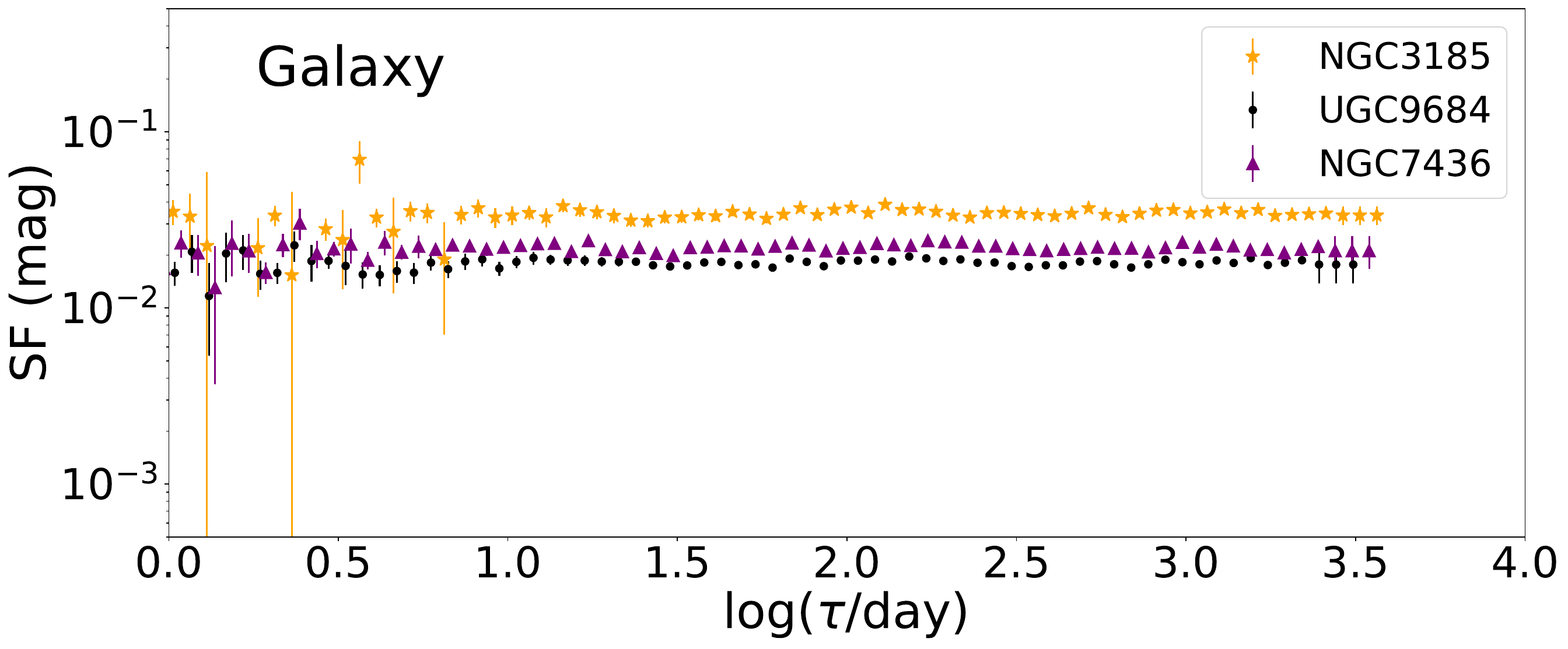}
\caption{Structure Function calculated using the ASAS-SN dataset. Top left: result for three individual Sy1s, Top Right: result for three individual Sy2s, Bottom: result for three individual galaxies. Structure functions of Sy1s rise at an earlier time than that of Sy2s.}\label{sf}
\end{figure}

We measure the structure functions of Sy1s, Sy2s and galaxies,  to further confirm the detection of Sy2 variability presented in the KS test using the ASAS-SN data set. While the structure function in general provides more detailed variability information for AGNs, for the low variability amplitude Sy2 sample, the interesting time scales are the very long scales where the structure function is approaching the variance of a light curve. Figure \ref{sf_stacked} shows the stacked structure functions of Sy1s, Sy2s, and galaxies. To exclude outliers in the galaxy light curves, we experimented with a cutoff between 5--6 $\sigma$ from the mean of the light curve.  We chose a 4.6$\sigma$ deviation from the mean cutoff, which resulted relative constant structure functions of galaxy light curves.  Modifying the cutoff value between 4.3--5 $\sigma$ does not significantly change the analysis results. The stacking was performed by the summation of individual structure functions for each group. Where the stacked structure function of Sy2s is significantly detected with a joint 8.5$\sigma$ detection beyond noise at long time scales. The stacked structure function of Sy1s shows a typical rising trend with a smooth break between a few hundreds to 1,000 days. 
The downturn at the longest timescales is potentially caused by the color-dependent variability effect due to combining the V and g-band ASAS-SN light curves, where the g filter is slightly bluer and has a slightly larger variability amplitude.
This color-dependent variability has a negligible effect for Sy2s and galaxies, since their variability amplitudes are very small or zero.
Another potential explanation is from the low sampling counts specifically at the longest timescale.
We also measured the significance of detections of individual Sy2 light curves. Among them 24 (63\%) are detected with $>10 \sigma$, eight (21\%) are detected between 3--10$\sigma$ and six (15\%) are detected with $<3 \sigma$. 
We show three significantly detected individual Sy2 structure functions in Figure~\ref{sf}, along with examples of individual structure functions of Sy1s and galaxies.
Compared to Sy1s, the detected structure functions of Sy2s rise at much longer time scales above noise. The structure functions of galaxies are consistent with noise. Both the KS test results on the normalized excess variance between the samples and structure function measurements demonstrate the detection of variability in Sy2s. We also examined the structure functions utilizing the TESS data set, and the Sy2s were consistent with galaxies, confirming again the results of the KS test that Sy2 variability is not detected in time scales shorter than the month long TESS window. 

\subsection{Estimation of the Scattering region}\label{Torus}
\begin{figure}[ht]
\centering
\includegraphics[width=0.3\textwidth]{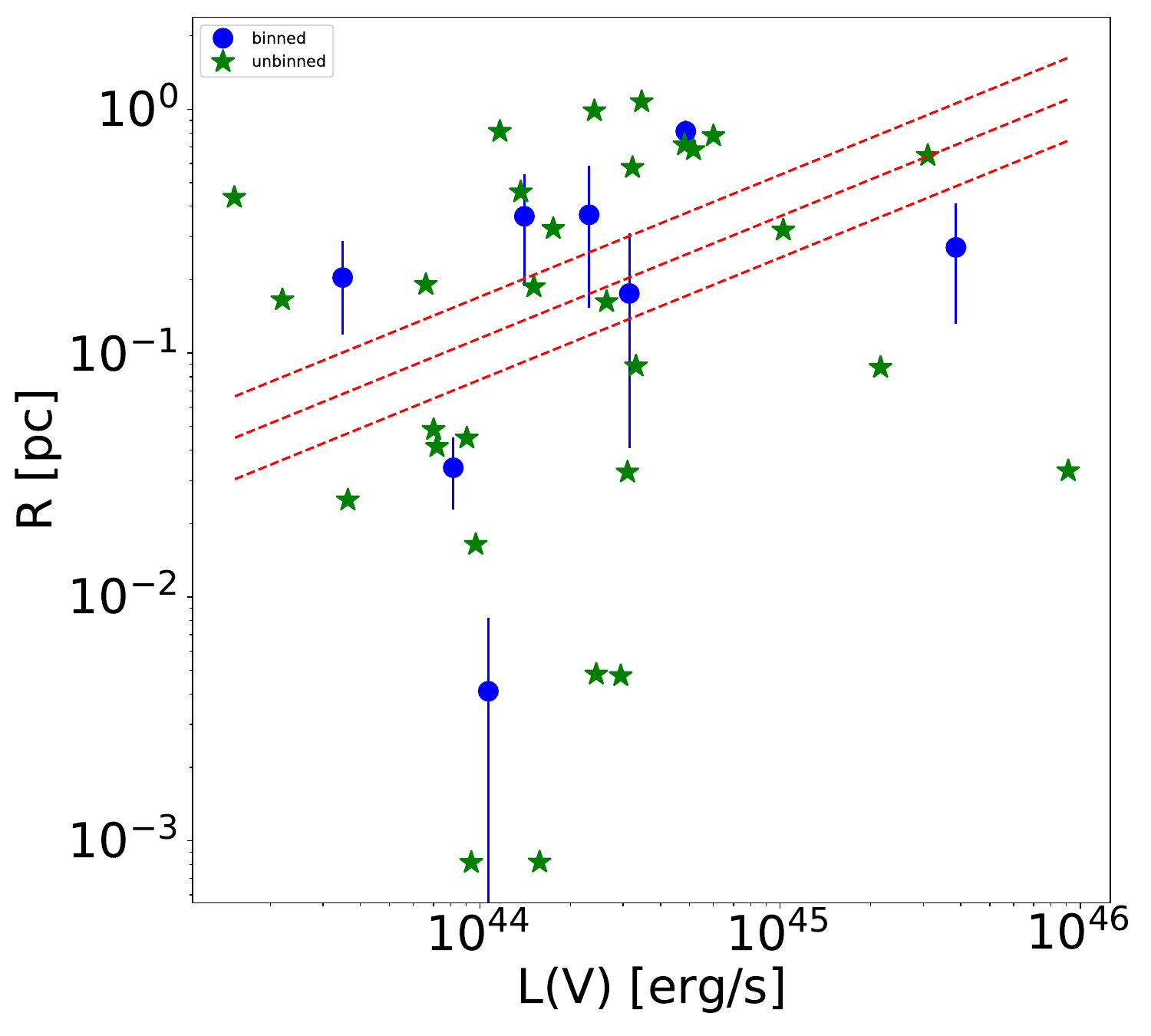}
\includegraphics[width=0.3\textwidth]{torus_size_relation_70.pdf}
\includegraphics[width=0.3\textwidth]{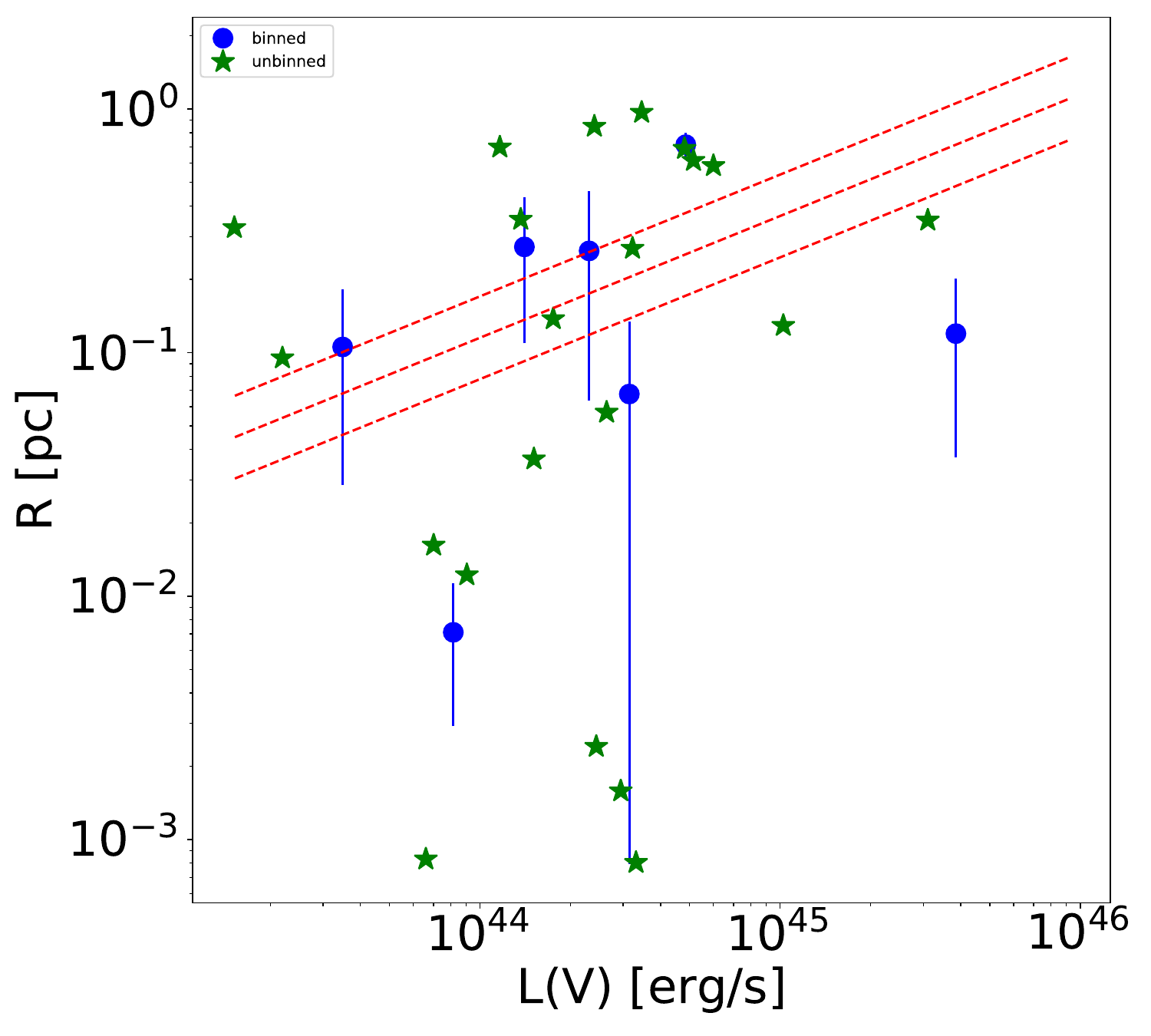}
\caption{Estimation of the size of the scattering region for our Sy1 sample, by smoothing Sy1s to a smoothing window from the scattering region to achieve  Sy2 variation amplitudes for different dilution percentages of host stellar light (Left: 60\% dilution, Middle: 70\% dilution, Right: 80\% dilution).
Showing both the binned in blue circles and unbinned in green stars. The dashed lines are the reverberation mapping torus
relation and its scatter, which is taken from \citet{Kishimoto_2007A&A...476..713K} with the upper and lower bounds from the scatter presented
in \citet{Yang_2020ApJ...900...58Y}.}\label{torus_size}
\end{figure}

Due to the obscuration of Sy2s by a dusty torus, in accordance with the unification model (\citealt{Antonucci_1993ARA&A..31..473A, Urry_1995PASP..107..803U}), the optical variability is suppressed and the Sy2 optical continuum is composed of the scattered emission and host contribution. One overly simple explanation is that the UV/optical light from the central engine is absorbed as it goes through the torus, which is reemitted in the infrared, resulting in the non-variable stellar light in the host galaxy to dominate the observed flux. However, this simple scenario does not explain the detected Sy2 variability presented in this paper. We present an enhanced interpretation based on the unification model, where a percentage of light from the central engine is scattered into our line-of-sight via a scattering region of a certain size with the added non-variable stellar light component. The variability of the central engine is diluted by two effects, the scattering and the non-variable host. In this scenario, the size of the scattering region can be determined by comparing the variability characteristics between Sy1s and 2s.
Here, we assume a scattering variability suppression model, where the unobscured light curve from the central engine is smoothed by a window of the scattering region size and a constant percentage of host stellar light is added to the flux of the smoothed light curve as illustrated in Figure \ref{interp} left. The suppression of the Sy1 light curves is performed by the 1D boxcar method, then the normalized excess variance is calculated with the added constant host flux. This is calculated after the light curve is suppressed by a series of window sizes, until the normalized excess variance matches that of the median value of the Sy2 normalized excess variance. The scattering region size is calculated using Equation \ref{equation4} below.

\begin{equation}\label{equation4}
R = \Delta t_{\textrm{rest}} \times c
\end{equation}
where $c$ is the speed of light and \(\Delta t_{\textrm{rest}}\) is the rest-frame smoothing window given by 
\begin{equation}\label{equation5}
\Delta t_{\textrm{rest}} = \frac{\Delta t_{\textrm{obs}}}{1+z}
\end{equation}
and \(\Delta t_{\textrm{obs}}\) is the observed window size for smoothing. We measured \(\Delta t_{\textrm{obs}}\) values ranging from roughly 250 days -- 1250 days.
We applied this method for all Sy1s using their ASAS-SN light curves and estimated the size of a scattering region to reduce their variability to typical values of Sy2s as if we changed the line-of-sight to a Sy2 perspective. This method was applied with different percentages (60\%, 70\%, 80\%) of optical stellar dilution added to the smoothed Sy1 light curves.
We display the results in Figure~\ref{torus_size}, the relation between the scattering region size and the V-band luminosity, alongside the torus size luminosity relation and its scatter presented in \citet{Kishimoto_2007A&A...476..713K} and \citet{Yang_2020ApJ...900...58Y} through dust reverberation mapping. The binned values were calculated using the average scattering region size in each bin, where each bin contained four data points, and the uncertainties were estimated using the standard error of the mean.
The scattering region size estimates from our study range $8.4\times10^{-4}$ to 0.99 pc with a median value of 0.13 pc. This is largely consistent with the torus luminosity relation, made evident with our binned measurements and an average dilution percentage of 70\%.  
This result constrains the size of the scattering region to be largely consistent with the size of the inner radius of the torus.

\section{Discussion and Conclusion} \label{disc}

The results presented from the KS test based on the ASAS-SN measurements in Figure \ref{KS} displays that Sy2s and galaxies exhibit a clear difference in variability, suggesting Sy2 variability is detected in the long time scale ASAS-SN light curves. Previously, a few Sy2s have been observed to potentially exhibit long time scale variations in the literature \citep{choi_2014ApJ...782...37C}; however, it was inconclusive whether the variability is distinguishable from noise and no clear explanation for the variability was discussed. Our analysis shows that long time scale variability of the AGN central engine longer than the light crossing time of the scattering region can pass through the scattering process. This is further seen from comparing the structure functions of Sy1s, Sy2s and galaxies, where the structure function at long time scales approaches the variance of the light curve, which includes all time scales. The structure function of a typical Sy2 rises above the constant noise at a later time scale than that of a typical Sy1.

We were able to estimate the size of the scattering region of the Sy1s in our sample utilizing the variability of the Sy2s and the concept of the unification model. This was done via a 1D boxcar smoothing of Sy1 light curves as an approximation to suppress their variability and match that of the typical Sy2. A fraction of stellar dilution was included in the Sy1 light curves to simulate the non-variable stellar light that dominates Sy2 light curves. This allows us to mimic the process, where the region smooths the unobscured photon packets from the central engine as the light scatters over the torus. Our method of calculating the size of the scattering region was compared to torus size estimations \citep{Yang_2020ApJ...900...58Y}, using reverberation mapping for their estimation. A test was performed on whether the scatter of the data could be reduced by incorporating a third parameter (i.e black hole mass or Eddington ratio). The incorporation of a third parameter, however, could not significantly reduce the $\chi^2$. Although a lot of our sources are showing a larger size compared to the torus size relation, our sizes are largely consistent with the inner radius of the torus. This scattering region is also responsible for the presence of broad lines in the polarized spectrum of Sy2s that exhibit an HBLR (\citealt{Heisler_1997Natur.385..700H,ichikawa_2015ApJ...803...57I}).

In our analysis we assumed a uniformly distributed torus. However, studies done on the distribution of dust within the torus suggest that it might not be uniform, as Silicate is observed in the MIR as an absorption feature in Sy2 spectra, acting as direct evidence of the existence of the dusty torus, while the expected emission feature from Sy1s is missing altogether \citep{Roche_1991MNRAS.248..606R}. The clumpy torus model was introduced as a means to explain the absence of the Silicate emission features in MIR spectra of Sy1s \citep{Nenkova_2002}. In the clumpy torus model, it is suggested that the dust is distributed in large clump like clouds within the torus (\citealt{Nenkova_2008,Nenkova_2008_2}). We plan to refine our method to include the clumpy torus model in future analysis work, where the covering fraction of the clumpy torus can potentially be inferred by modeling the leaked short time scale variability from the central engine, where our current TESS measurements may suggest a small amount of leaked variability in Sy2s compared to galaxy light curves.

\begin{acknowledgments}
We thank the anonymous referee for helpful comments.
N. K., X. D., and H. Y.
would like to acknowledge NASA funds 80NSSC22K0488, 80NSSC23K0379 and NSF fund AAG2307802. We thank
Las Cumbres Observatory and its staff for their continued support of ASAS-SN. ASAS-SN is funded by Gordon and
Betty Moore Foundation grants GBMF5490 and GBMF10501 and the Alfred P. Sloan Foundation grant G-2021-14192.
\end{acknowledgments}

\begin{contribution}

\end{contribution}

\facilities{ASAS-SN, TESS, \textit{Swift}}

\bibliography{bibliography}{}
\bibliographystyle{aasjournalv7}

\end{document}